\documentclass[prd,showpacs,amsmath,showkeys,twocolumn,floatfix]{revtex4} 
\usepackage[]{epsfig,dcolumn}
\usepackage{graphicx}
\usepackage{epstopdf}
\usepackage{bm}

\def\x{{\bf x}}
\def\y{{\bf y}}
\def\z{{\bf z}}
\def\k{{\bf k}}
\def\q{{\bf q}}
\def\p{{\bf p}}

\def\A{{\bf A}}
\def\B{{\bf B}}

\def\D{{\bf D}}
\def\R{{\bf R}}
\def\E{{\bf E}}
\def\L{{\bf L}}
\def\l{{\bf l}}
\def\R{{\bf R}}
\def\r{{\bf r}}
\def\na{\bm{\nabla}}

\def\lsim{\mathrel{\rlap{\lower4pt\hbox{\hskip1pt$\sim$}}
    \raise1pt\hbox{$<$}}}
\def\gsim{\mathrel{\rlap{\lower4pt\hbox{\hskip1pt$\sim$}}
    \raise1pt\hbox{$>$}}}

\newcommand{\eref}[1]{Eq.~(\ref{#1})}
\begin{document}


\title{ Chromo-Electric flux tubes } 

\author{ Patrick O.~Bowman and Adam P.~Szczepaniak }
\affiliation{ Physics Department and Nuclear Theory Center \\
Indiana University, Bloomington, Indiana 47405 }

\date{\today}

\begin{abstract}
The profiles of the chromo-electric field generated by static quark-antiquark,
$Q{\bar Q}$ and three-quark, $QQQ$ sources are calculated in Coulomb gauge.  
Using a variational ansatz for the ground state, we show that a flux tube-like 
structure emerges and combines to the ``Y''-shape field profile for three
static quarks.  The properties of the chromo-electric field are, however, not 
expected to be the same as those of the full action density or the Wilson line 
and the differences are discussed. 
\end{abstract}


\maketitle
\section{Introduction} 

An intuitive picture of quark-gluon dynamics emerges in the 
 Coulomb gauge, $\na\cdot \A^a = 0$~\cite{clee,zw1,zw2}.
 In this case QCD is represented as a many-body system of strongly 
 interacting physical quarks, antiquarks and gluons. In particular the gluon
 degrees of freedom have only the two transverse polarizations and in the 
 non-interacting limit reduce to the physical massless plane wave
 states. In the interacting theory gluonic states, just like any other 
 colored objects, are expected to be non-propagating, {\it i.e.} confined 
 on the hadronic scale. The non-propagating nature of colored states 
 follows from the infrared enhanced dispersion relations which can be 
 set up in the Coulomb gauge~\cite{zw2,as1,as2,as3}. 

In the Coulomb gauge the $A^0$ component of the 4-vector potential 
 results in an instantaneous interaction (potential) between color
 charges. Unlike QED, where the corresponding potential is a function only
 of the relative distance between the electric charges, in QCD it
 is a functional of the  transverse gluon components, ${\bf A}$~\cite{clee}. 
 Thus the numerical value of the  potential cannot be obtained without
 knowing the correct wave functional of the state and its dependence
 on the gluon coordinates.  So in QCD the chromo-electric field is 
 expected to be non-local and to depend on the global distribution of 
 charges, which set up the gluon wave functional. 

Even though the exact solution to the general many-body problem is unavailable 
 it is often possible to obtain good approximations if the dominant
 correlations can be identified. In Coulomb gauge QCD (in the
 Schr\"odinger field representation) the domain of the transverse
 gluon field, $\A$ is bounded and non-flat, and is referred to as the 
 Gribov region. It is expected that the strong interaction between
 static charges originates from the long-range modes near the boundary
 of the Gribov region, the so called Gribov horizon. For example 
 it has been recently shown that center vortices, when transformed to
 the Coulomb gauge, indeed reside on the Gribov horizon~\cite{goz}.  

The curvature of the Gribov region contributes to matrix elements via
 the functional measure determined by the determinant of the
 Faddeev-Popov operator. This determinant prevents analytical
 calculations of functional integrals, however it has been shown  
 that its effect can be approximated by imposing appropriate boundary
 conditions on the gluon wave functional~\cite{as4,hr}. 
 This wave functional is in
 turn constrained by minimizing the expectation value of the energy
 density which leads to a set of coupled self-consistent Dyson
 equations~\cite{as1,sw}.  Once the wave functional is determined it is 
 possible to calculate the distribution of the chromo-electric field
 in the system. This is the main subject of this paper.  

In the following we study the chromo-electric field in the presence of
 the static quark-antiquark and three-quark systems, prototypes for a
 meson and a baryon respectively.  Recent lattice computations
 indicate that the gluonic field near the static $Q-{\bar Q}$ state
 forms flux tubes. There are also indications that for the $QQQ$ state 
 the fields arrange in the so called ``Y''-shape~\cite{tak,latY}, although
 some work supports the ``$\Delta$''-shape~\cite{latD}.  String-like behavior
 has been observed in the chromo-electric field in Ref.~\cite{sho}
 and the ``Y''-shape interaction advocated in Ref.~\cite{kuzsim}.  A recent
 reevaluation of the center-vortex model also supports the 
 ``Y''-shape~\cite{corn}.  
 
 In the following section we summarize the relevant elements of the Coulomb 
 gauge formalism and discuss the approximations used.  This is followed by 
 numerical results and outlook of future studies. 
 There is a fundamental difference between lattice gauge flux tubes 
 corresponding to the distribution for the action density and the 
 chromo-electric field profiles.  
 In the context of the potential energy of the sources, this difference was 
 emphasized in Zwanziger, Greensite and Olejnik~\cite{goz, zwancon}. We discuss
 those in Section.~IV.

\section{Chromo-electric Coulomb field in the presence of static charges} 

\subsection{The Coulomb gauge Hamiltonian}
The Yang-Mills Coulomb gauge Hamiltonian in the Schr\"odinger
representation, $H=H(\Pi, A)$ is given by 
\begin{equation}
H  =  \frac{1}{2} \int d\x \left[ \bm{\Pi}^a(\x) \cdot \bm{\Pi}^a(\x) 
 +    \B^a(\x)\cdot \B^a(\x) \right] + {\hat V}_C. \label{h}
\end{equation}
 The gluon field satisfies the Coulomb gauge condition, 
 $\na \cdot \A^a(\x)=0$,  for all color components 
 $a=1\cdots N_c^2-1$.  The conjugate momenta, $\bm{\Pi}^a(\x)=  
   -i\partial/\partial {\A^a(\x)}$ obey the canonical commutation
 relation,   $[\Pi^{i,a}(\x), A^{j,b}(\y)] = -i\delta_{ab} 
  \delta^{ij}_T(\na_\x)\delta(\x-\y)$, with  $\delta^{ij}_T(\na) =
 \delta_{ij} - \nabla_i \nabla_i/\na^2$. 
 The canonical momenta also correspond to the negative of 
 the transverse component of the  chromo-electric field, $\bm{\Pi}^a(\x) = - 
 \E^a_T(\x)$, $\na \cdot \E^a_T = 0$.
 The chromo-magnetic field, $\B$ contains 
 linear and quadratic terms in $\A$. It will also be convenient to 
 transform to the momentum space components of the fields by 
 \begin{equation}
\A^a(\k) = \int d\x\A^a(\x) e^{-i\k\cdot x},
\end{equation}
and similarly for $\bm{\Pi}^a(\k)$. 
The Coulomb potential ${\hat V}_C$ may be expressed in terms of the
longitudinal component of the chromo-electric field, 
\begin{equation}
{\hat  V}_C = \frac{1}{2}\int d\x {\bf E}^a(\x) {\bf E}^a(\x),
\end{equation}
with
\begin{equation}
{\bf E}^a(\x) = \int d\y d\z { \frac{\x - \y}{4\pi|\x - \y|^3}} 
 \left[ \frac{g}{1 - \lambda}\right]^{ab}_{\y,\z} \rho^b(\z). \label{e}
\end{equation}
Here $(1-\lambda)$ is the Faddeev-Popov (FP) operator which in 
 the configuration-color space is determined by, 
\begin{equation}
[\lambda]^{ab}_{\x,\y} = \int {\frac{d\p}{(2\pi)^3}} {\frac{d\q}{(2\pi)^3}} 
   e^{i\p\cdot\x} e^{-i\q\cdot\y} 
\lambda_{ab}(\p,\q),
\end{equation}
where
\begin{equation}
\lambda_{ab}(\p,\q) = ig f_{acb} {\frac{\A^c(\p-\q)\cdot \q}{\q^2}},
\end{equation}
 $f$ are the $SU(N_c)$ structure constants, and $g$ is the bare coupling.  
  In \eref{e}, $\rho$ is the color charge density  given by
\begin{equation}
\rho^a(\x) = \psi^{\dag}(\x)T^a \psi(\x) + f_{abc} \A^b(\x)\cdot 
\bm{\Pi}^c(\x),
\end{equation}
with the two terms representing the quark and the gluon contribution, 
 respectively; the former is replaced by a $c$-number for static
 quarks.  Without light flavors there is no other dependence on the 
 quark degrees of freedom. The energy of the  static $Q{\bar Q}$ or $QQQ$
 systems measured with respect to the state with no sources 
 is thus given by the Coulomb term and is determined  by the  expectation value of the  longitudinal component of the chromo-electric field.
 
It is the dependence of the chromo-electric field and the Coulomb
interaction on the static vector potential (through $\lambda$) that produces
the differences between QCD and QED. In QED the 
 kernel in the bracket in \eref{e} reduces to $[\cdots] \to
 \delta(\y-\z)$ and the Abelian expression for the electric 
 field emerges. In QCD the chromo-electric field
 and  the Coulomb potential are enhanced due to 
 long-wavelength transverse gluon modes on the Gribov horizon where
 the FP operator vanishes.  The combination of two effects on the
 Gribov horizon:  enhancement of 
 $(1 - \lambda)^{-1}$ in the longitudinal electric field and vanishing
 of the functional norm, which is  proportional to $\det(1-\lambda)$,  
 leads to finite, albeit large, expectation values of the static
 interaction between color  charges.  In \eref{h} we have omitted
 the FP measure since, as mentioned earlier in Ref.~\cite{as4}, 
 its effect can be approximately accounted for by imposing specific
 boundary conditions on the ground state wave functional. 
 
Since the chromo-electric field depends on the distribution of the
transverse  vector potential it is necessary to know the wave
functional of  the system. A self-consistent 
 variational ansatz can be chosen in a Gaussian form, 

\begin{equation}
\Psi[A] = \exp\left( - \frac{1}{2}
 \int {\frac{d\p}{(2\pi)^3}} \omega(p)
  \A^a(\p)\cdot \A^a(-\p) \right). \label{varia}
\end{equation}
The parameter $\omega(p)$ ($p\equiv |\p|$) is determined by minimizing
the  expectation value of the  energy density of the vacuum ({\it i.e.} 
without sources). 
The boundary condition
referred to above corresponds to setting $\omega(0) \equiv \mu$ to be
finite, which plays the role of $\Lambda_{QCD}$, {\it i.e} it controls the
position of the Landau pole. Minimizing the energy density of the vacuum
leads to a set of coupled self-consistent integral equations: 
 one for $\omega$, one for the expectation  value of the inverse of the FP
 operator,  $d(p)$,
\begin{multline}
(2\pi)^3 \delta(\p-\q)\delta_{ab} d(p)
 \equiv \\
 \int d\x d\y e^{-i\p\cdot\x} e^{i\q\cdot\y} 
 \langle \Psi| \left[ \frac{g}{1 - \lambda}\right]^{ab}_{\x,\y}
 |\Psi \rangle / {\langle \Psi|\Psi \rangle },
\label{d}
\end{multline}
and one for the expectation value of the square of the inverse of the FP
operator, which appears in the matrix elements of $V_C$,
\begin{multline}
(2\pi)^3 \delta(\k-\q)\delta_{ab} f(p)d^2(p) 
 \equiv \\
 \int d\x d\y e^{-i\p\cdot\x} e^{i\q\cdot\y} 
 \langle \Psi| \left[ \left(\frac{g}{1 - \lambda}\right)^2
 \right]^{ab}_{\x,\y}
 |\Psi \rangle / {\langle \Psi|\Psi \rangle }.
\label{ff}
\end{multline}
The approximation $f=1$ ignores the dispersion in the expectation 
 value of the inverse of the FP operator,  
\begin{equation}
\Bigg\langle \left[ \frac{g}{1 - \lambda} \right]^2 \Bigg\rangle 
 \to  \Bigg\langle \frac{g}{1 - \lambda} \Bigg\rangle^2. \label{disp}
\end{equation}
This approximation has been extensively used, {\it e.g.} in 
Refs.~\cite{zw1,zwan}. 
The three Dyson equations were analyzed in Ref.~\cite{as1} where it was
found that the solution of $\omega$ can be well approximated by the 
simple function $\omega(p) = \theta( \mu - p) \mu + \theta( p -\mu)
p$. The renormalization scale $\mu$, being the only parameter in the
theory, can constrained by the long range part of the Coulomb kernel 
 $\langle V_C \rangle \propto fd^2$. We will discuss this more in the
 subsection below.  The low momentum, $p<\mu$ dependence of  $d(p)$ and of the
 Coulomb  potential $V_C(p) = f(p)d(p)^2$ is well approximated by a power-law, 
\begin{equation}
d(p) = d(\mu) \left( \frac{\mu}{p}\right)^\alpha, f(p) = f(\mu) 
\left( \frac{\mu}{p}\right)^\beta \label{df}
\end{equation}
 with $\alpha \sim 0.5$ and $\beta \sim 1$. The exponents are bounded by 
$ 2\alpha + \beta \le 2$ and the upper limit corresponds to the
linearly  rising confining potential. 
 At large momentum, $p >> \mu$, as expected 
 from asymptotic freedom, both $d$ and $f$ are proportional to 
 $1/\log^\gamma(p)$, with $\gamma = O(1)$. 
 Adding static sources does not modify the parameters of the vacuum
 gluon distribution, {\it e.g.} $\omega(p)$. This is because the
 vacuum energy is an  extensive quantity while sources contribute a 
 finite amount to the total energy. Thus we can use the three 
 functions $\omega$, $d$ and $f$ calculated in the absence of the
 sources to compute the expectation value of the 
 chromo-electric field in the presence of static sources.  The 
 ansantz state obtained by applying quark sources to the variational 
 vacuum of Eq.~(\ref{varia}) does not, however optimize the state with 
 sources.

\subsection{ The field lines in the $Q{\bar Q}$ and $QQQ$ systems }  

 For a quark and an antiquark at positions $\x_q \equiv \R/2 = 
  R{\hat \z}/2$ and  $\x_{\bar q} = -\R/2 = -R{\hat \z}/2$, respectively, 
 and the gluon field 
 distributed according to $\Psi[\A]$,  the expectation value of the 
 square of the magnitude of the chromo-electric field measured at
  position $\x$ is given by

\begin{multline}
\langle \E^2(\x,\R) \rangle   =  { \frac{C_F}{(4\pi)^2}}
  \sum_{\z_1=\pm \R/2} \sum_{\z_2 =\pm \R/2} \pm 
 \int d\y_1 d\y_2 \\
\times{  \frac{ (\x - \y_1)\cdot (\x-\y_2)}{|\x - \y_1|^3
 |\x - \y_2|^3}} E(\z_1,\y_1;\z_2,\y_2), \label{qq}
\end{multline}
where the $+ (-)$ sign is for the $\z_1 =(\ne) \z_2$ contributions,  
 and 
\begin{equation}
 E(\z_1,\y_1;\z_2,\y_2) 
 \equiv \frac{\langle \Psi | 
\left[ \frac{g}{1 - \lambda}\right]_{\z_1,\y_1}
  \left[ \frac{g}{1 - \lambda}\right]_{\y_2,\z_2}|\Psi  \rangle }
{ \langle \Psi|\Psi \rangle }. \label{ee}
 \end{equation}
 The color factors leading to $C_F$ can be extracted from 
  the expectation value in \eref{ee} (the ground state expectation
  value of the inverse of two FP operators is 
  an  identity in the adjoint representation).   
 In the Abelian limit, $E(\z_1 \cdots \y_2) \to \delta(\y_1 -
\z_1)\delta(\y_2 - \z_2)$ and Eq.~(\ref{qq}) gives the dipole field
 distribution, $\langle \E^2 \rangle_{QED}$. One should note that
 \eref{qq} contains the two self energies. These self energies are
 necessary to produce the correct asymptotic behavior at $x >>
 R$ for charge-neutral systems, (in QED and QCD) {\it i.e} $\E^2$
 has to fall-off at least as $1/\x^4$ at large distances from the sources.   

The infrared, $|\x| \sim |\R| >> 1/\mu$  enhancement in
 QCD arises from the expectation value of the inverse of the 
 FP operator. If $\langle \E^2(\x,\R)  \rangle$ is integrated over
 $\x$  one obtains the expectation value of the Coulomb energy of the 
 $Q{\bar Q}$ source.  The mutual interaction energy is given by, 
\begin{align}
V_C(\R) &= {1\over 2}\int d\x \langle \E^2(\x,\R) \rangle \nonumber \\ 
&\hspace{-5mm}=  -C_F {{\langle \Psi | 
\left[ {g\over {1 - \lambda}} \left(-{1\over \na^2}\right)
       {g\over {1 - \lambda}}\right]_{{\R\over 2},-{\R\over 2}}
 |\Psi  \rangle }
/{ \langle \Psi|\Psi \rangle } 
}, \nonumber \\ 
&\hspace{-5mm} = -C_F \int {{d\p}\over {(2\pi)^3}} {{d^2(p) f(p)}\over {p^2}}
 e^{i\p\cdot \R},
\label{vc} 
\end{align}
and the net self-energy contribution is, 
\begin{align}
\Sigma &=  C_F {{\langle \Psi | 
\left[ {g\over {1 - \lambda}} \left(-{1\over \na^2}\right)
       {g\over {1 - \lambda}}\right]_{\pm {\R\over 2},\pm {\R\over 2}}
 |\Psi  \rangle }
/{ \langle \Psi|\Psi \rangle } }, \nonumber \\
 &=  C_F \int {{d\p} \over {(2\pi)^3}} {{d^2(p) f(p)}\over {p^2}}.
 \label{sigma} 
\end{align}
 In lattice simulations it has been shown~\cite{Greensite} 
  that the Coulomb energy and the phenomenological static $Q\bar Q$ potential
 obtained from the Wilson loop are different. In particular it was
  found that the Coulomb potential string tension is about three times
larger than the phenomenological string tension. This is 
 in agreement with the ``no confinement without Coulomb Confinement'' 
 scenario discussed by Zwanziger~\cite{zwancon}.  
 It is simple to understand the origin
  of the difference. Even if $|\Psi[\A]\rangle $ were the true vacuum
  state (without sources) of the Coulomb gauge QCD Hamiltonian 
 (here we approximate it by a variational ansatz) the state 
 $|Q{\bar Q},R\rangle \equiv  Q(\R/2){\bar Q}(-\R/2) |\Psi[A]\rangle$
  is no longer an eigenstate. For example ${\hat V}_C$ acting on $|Q{\bar
  Q},R\rangle$ excites any number of gluons and couples them to the
  quark sources. The Coulomb energy was defined as the expectation value, 
 $V_C$ in $|Q{\bar Q},R\rangle$ minus the vacuum energy and it is
  therefore different from the phenomenological static 
 potential energy which corresponds to the total energy (measured with 
 respect to the vacuum) of the true eigenstate of the Hamiltonian 
 with a $Q{\bar Q}$ pair. If one defines~\cite{goz}  
 \begin{align} 
G(R,T) &\equiv  \langle Q{\bar Q},R|e^{-(H-E_0)T}|Q{\bar Q},R\rangle 
 \nonumber \\
& = \sum_n |\langle Q {\bar Q},R,n|Q{\bar Q},R\rangle|^2 e^{-(E_n -
 E_0) T},
\end{align}
then the Coulomb potential on the lattice can be calculated from 
\begin{equation}
V_C(R) = \lim_{T=0} -{d\over {dT}} \log(G(R,T)),
\end{equation}
and the phenomenological potential from 
\begin{equation}
V(R) = \lim_{T=\infty} -{d\over {dT}} \log(G(R,T)).
 \end{equation}
 Thus one should be comparing $V_C(R)$ in \eref{vc} to the lattice
 Coulomb potential and not to the phenomenological potential obtained
 from the Wilson loop. Finally, one could try to optimize the state with 
 sources, {\it e.g.} by adding gluonic components.  In this case terms in the 
 Hamiltonian beyond the Coulomb term would contribute to the energy of the 
 system and one could compare with the true (Wilson loop) static energy.  
  In our previous studies, where we extracted
 numerical values for $\mu$ and the critical exponents
 $\alpha,\beta,\gamma$ ({\it cf.} \eref{df}) we have instead compared
 $V_C$ to the phenomenological, Wilson potential~\cite{as1}. 
 In what follows we will use the larger value of the string
 tension, to be in agreement with Ref.~\cite{goz}.  

 If the two exponents $\alpha$ and $\beta$, which determine the
 infrared behavior of $d(p)$ and $f(p)$ respectively, 
 satisfy  $2\alpha + \beta > 2$, then the self energy in \eref{sigma} 
 is divergent and so is the {\it rhs} of \eref{vc}. 
 This reflects the long-range
 behavior of the effective confining potential generated by
 self-interactions between the gluons that make up the Coulomb operator.  
 For the colorless $Q\bar Q$ system the total energy which is the 
 sum of $V_C$ and $\Sigma$, is finite as it should be. For
 a colored system, {\it e.g.} a quark-quark source, the sign of $V_C$
 changes, there is no cancellation between the infrared singularities, and
  in the confined phase the system would be un-physical with infinite energy. 
 The integral determining the self energy also becomes divergent in the
 UV, since for $p\to \infty$ the product $d^2(p)f(p)$ only falls-off 
 logarithmically. 
 Modulo these logarithmic corrections this UV divergence is the same as in the
 Abelian theory and can be removed by renormalizing the quark charge. 
 
It follows from translational invariance of the matrix element  
 in \eref{ee}, that $E$ depends only on the relative coordinates, 
 $\z_1 - \y_1$ and $\z_2 - \y_2$. We therefore introduce the momentum space 
 representation, 
\begin{eqnarray}
E(\z_1,\y_1;\z_2,\y_2) & = & \int {{d\p} \over {(2\pi)^3}} 
  {{d\q} \over {(2\pi)^3}} e^{i\p \cdot (\z_1 - \y_1)  
 -i \q\cdot (\z_2 - \y_2) } \nonumber \\ 
& &  \qquad\times d(p) d(q) E(\p;\q), \label{ft}
\end{eqnarray}
 and define $F_\L(\l) \equiv E(\l+\L/2; \l-\L/2)$ with 
$\l \equiv (\p + \q)/2$ and  $\L  \equiv \p - \q$. 
 The Dyson equation for $F$ can be derived in the
 rainbow-ladder approximation which, as shown in Ref.~\cite{as1,as2}, 
 sums up the dominant infrared and ultra-violet contributions to the 
 expectation value of the inverse of two FP operators,  
\begin{widetext}
\begin{equation}
F_\L(\l)
 = 1 + N_c \int {{d\k}\over {(2\pi)^3 }} 
{{\left[ (\k-\L/2)\delta_T(\k + \l)(\k+\L/2) \right] } 
 \over  {2\omega(\k + \l)}}
 {{d(\k - \L/2)}\over {(\k - \L/2)^2}}
 {{d(\k + \L/2)}\over {(\k + \L/2)^2}} F_\L(\k), \label{F}
\end{equation}
\end{widetext}
 It follows from \eref{ft}  that $\L$ and $\l$ are conjugate to 
 the {\it center of mass},  $\R \equiv [ (\z_1 - \y_1) + (\z_2 - \y_2)
 ]/2$ and  the {\it relative}, $\r \equiv [ (\z_1 - \y_1) - (\z_2 - \y_2)
 ]$ coordinate respectively.  
 The Dyson equation for $F_L$ is UV divergent if for $p/\mu >> 1$, and
   $d(p) \ge \log^{1/2}(p^2)$. This divergence can be removed 
   by the Coulomb operator renormalization  constant. 
 The renormalized equation is obtained from the once-subtracted
   equation $F_\L(\l) - F_{\L_0}(\l_0)$. For example, if the
   subtraction is chosen at $|\l_0|  = \mu$ and $\L_0 = {\bf 0}$, the 
   renormalized coupling $F_0(\mu)$ can be fixed
   from the Coulomb potential. After integrating \eref{qq} (over
   $\x$) one obtains $\delta(\y_1 -
   \y_2)$ multiplying $E(\z_1, \cdots, \y_2)$. Therefore, it
   follows from \eref{ft} that $V_C(\R)$ is determined by 
 $F_0(\l)$ and $F_0(\l) = F_0(l) = f(l)$ with  $f$ defined in \eref{ff}. 

   In \eref{F} $\L$ is a parameter, 
 {\it i.e.} the Dyson equation does not involve self-consistency in
 $\L$. We have just shown that as $\L \to {\bf 0}$, $F_\L(\l)$ has a finite
 limit: it is given by $f$.  For large $L=|\L|$ ($L/\mu >> 1$), 
   due to asymptotic freedom, $F_\L$ is expected to vanish
   logarithmically, $F_\L \to d^2(L) \propto 1/\log(L^2)$.  
 We do not attempt here to
   solve \eref{F}, instead we use a simple interpolation formula
   between the $L=0$ and $L\to \infty$ limits, 
\begin{widetext}
\begin{equation}
 F_\L(\l) = 
 f(\l) \theta(\mu - |\l|) \theta(\mu -|{\L\over 2}|) + \left[ 1 - 
 \theta(\mu - |\l|)\theta(\mu - |{\L\over 2}|) \right]  
 \sim f\left( {\p + \q}\over 2 \right) \theta(\mu - p) \theta( \mu - q) 
  + \left[ 1 - \theta(p) \theta(q) \right],  \label{f}
\end{equation}  
\end{widetext}
{\it i.e.} in the term in the bracket we ignore the short distance logarithmic corrections.  It is easy to show that if logarithmic corrections
  are ignored then the short-range, $p,q > \mu$ contribution to the energy 
 density is the same as in the Abelian case. 
 Since we are mainly interested in the long range behavior
 of the chromo-electric field, in the following 
  we shall ignore contributions from the region $p,q>\mu$ all together. 
 In the long-range approximation, $x, R >> 1/\mu$ the expectation 
 value of $\E^2$ is then given by,  
\begin{widetext}
\begin{equation}
\langle \E^2(\x,\R) \rangle = 
 {{C_F}\over {(4\pi)^2}}  
\sum_{ij=1}^2 \xi^{Q\bar Q}_{ij} \int d\r f_L(r) 
  {{\z_i - \x - \r/2} \over {|\z_i - \x - \r/2|}}
\cdot   {{\z_j - \x + \r/2} \over {|\z_j - \x + \r/2|}}  
 d'_L(\z_i - \x - \r/2) d'_L(\z_j - \x + \r/2).
  \label{el}
\end{equation}
\end{widetext}
$\xi^{Q\bar Q}_{ij} = 1$ for $i=j$ and $-1$ for $i\ne j$, 
 $\z_{1,(2)} = (-)\R/2$, 
\begin{equation}
d'_L(r) \equiv  {2\over {\pi}}\int^\mu_0 p dp j_1(rp) d(p), \label{dpl}
\end{equation}
is the derivative of $d_L$ w.r.t.\ $r$,
\begin{equation}
f_L(r) = {1\over {2\pi^2}} \int^\mu_0 dp p^2 f(p) j_0(pr), \label{fl}
\end{equation}
and $j_0, j_1$ are Bessel's functions.
We note that the expression in
 \eref{el} is not necessarily positive. In the limit $f(p)=1$,
 the matrix element of the square of the inverse of the FP operator 
 is approximated  by the square of  matrix elements (cf. \eref{disp})
 and $\langle \E^2\rangle$ becomes positive.

  The expression for $\langle \E^2\rangle$ for the three quark system
 is derived by taking the expectation value of the Coulomb operator,
 ${\hat V}_C$ in a color-singlet state $\epsilon_{ijk}Q_i(\z_1) Q_j(\z_2)
 Q_k(\z_3) |\Psi[\A]\rangle$, which gives 

\begin{widetext}
\begin{equation}
\langle \E^2(\x,\R_i) \rangle = {{C_F}\over {(4\pi)^2}} 
  \sum_{ij=1}^3 \xi^{QQQ}_{ij} \int d\r f_L(r) {{ \z_i - \x - \r/2 } \over 
 { |\z_i - \x - \r/2| } }
\cdot
{{ \z_j - \x + \r/2 } \over 
 { |\z_j - \x + \r/2| }} d'_L(\z_i - \x - \r/2) 
 d'_L(\z_j - \x + \r/2) 
\end{equation}
\end{widetext}
where $\xi^{QQQ}_{ij} = 1$ if $i = j$ and $\xi^{QQQ}_{ij} = -1/2$ if
$i \ne j$. 
We note that the energy density for the $QQQ$ system comes
from two-body correlations between the $QQ$ pairs.

\section{ Numerical results } 
We first consider the simple approximation to the expectation value of
the Coulomb kernel of \eref{disp} in which $f(p)=1$. If one wishes to
have the confining potential grow linearly at large distances then it is 
necessary to set $\alpha = 1$, {\it i.e.} $d(p) \propto \mu/p$ for
$p/\mu < 1$. In this case, assuming that the long-range
behavior of the potential is of the form $V_C(r) = b_C r$, we obtain
from \eref{vc},
\begin{equation}
b_C = C_F d^2(\mu)\mu^2/(8\pi). \label{bc}
\end{equation}
We use the Coulomb string tension $b_C  = 0.6 \mbox{ GeV}^2$.
For the $Q\bar Q$ system the long-range contribution to the electric fields is
then given by, 
\begin{multline}
\langle \E^2(\x,\R) \rangle = 
  {{2b_C}\over {\pi^3}}
\biggl[
  {{(\R/2 - \x)} \over {|\R/2 -
  \x|^2}}\left(1-j_0(\mu|\R/2-\x|)\right) \\
 + (\x \to -\x) \biggr]^2. \label{elnof}
\end{multline}

 \begin{figure}
 \includegraphics[width=3.in]{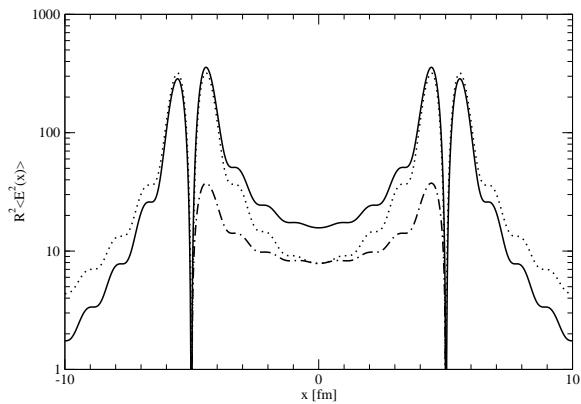}
 \caption{\label{fig1} $R^2 \langle \E^2(x) \rangle$ in units of 
  $2b_C/\pi^3(\hbar c)^2 $ as a function of the distance $x$ along the 
  $Q{\bar Q}$ axis. We employ the $f(p) = 1$ approximation.  The quark and the 
  antiquark are located at $R/2=5\mbox{ fm}$
  and $-R/2 = -5\mbox{ fm}$ respectively. The renormalization scale
  $\mu=1.1\mbox{ GeV}$ is calculated from \eref{bc} using 
  $d(\mu) = 3.5$  from
  Ref.~\cite{as1}. The dashed line is the
  contribution from the two self energies, the dash-dotted line
  represents mutual interactions and the solid line is the total.}
\end{figure}

  In Fig.~1  we show the Coulomb energy density as a function of
 position on the $Q{\bar Q}$ axis, $\x = x\hat{\R}$, for $R=|\R|=10\mbox{ fm}$.
  The small oscillations come from the sharp-cutoff introduced
 by the $\theta$-functions in \eref{f} which produces
 the Bessel's functions in \eref{elnof}. For a smooth
 cutoff, {\it e.g.} with $\theta(\mu - p) \to \exp(-p/\mu)$ in
 \eref{elnof}  one should 
 replace $1 - j_0(\mu|\R/2-\x|)$ by $1 - \arctan(\mu|\R/2-\x|)/\mu|\R/2-\x|$. 
 The cut-off is also responsible for the rapid variations near the quark positions, $\x = \pm R/2$. 
 
   We note that for large separations between the quarks, $R >> 1/\mu$ 
 and $x << R$, the Coulomb energy density behaves as expected from 
 dimensional analysis, 
\begin{equation}
\langle \E^2(\x,R\mu\to \infty) \rangle \to {{32 b_C}\over {\pi^3R^2}}.
 \label{short}
\end{equation}
 which is consistent with linear confinement, {\it i.e.} if $\langle \E^2(\x,R\mu\to \infty) \rangle$ is integrated over $\x$ in the region $|\x|<R$ on obtains $V_C(R)  \propto R$. 
 
  At large distances 
 $x >> R >> 1/\mu$ we obtain 
\begin{equation}
\langle \E^2(|\x|/R \to \infty,R\mu\to \infty) \rangle \to 
 {{2 b_C R^2}\over {\pi^3\x^4}}. \label{inf} 
\end{equation}
If there were a finite correlation length one would expect 
$\langle \E^2(|\x|/R \to \infty,R\mu\to \infty)\rangle$ to fall-off 
exponentially with $|\x|$~\cite{DDS}  and not as a power-law. The power-law 
behavior obtained in Eq.~(\ref{inf}) is again related to the difference 
between the $|Q{\bar Q}, R\rangle$ state used here, which is built by adding 
quark sources to the vacuum and the true ground state of the $Q{\bar Q}$ 
system as discussed in Sec.~IIA. In other words the profile of the 
chromo-electric field distribution for such a state is not expected to agree 
with the profile of the flux-tube or action density. 
    To illustrate this difference, in Fig.~2 we plot the energy density as a 
function of the magnitude of the distance transverse to the $Q\bar Q$ axis, 
$x_\perp = |\x_\perp|$, $\R\cdot \x = \R\cdot \x_\perp = 0$.

 \begin{figure}
 \includegraphics[width=3.in]{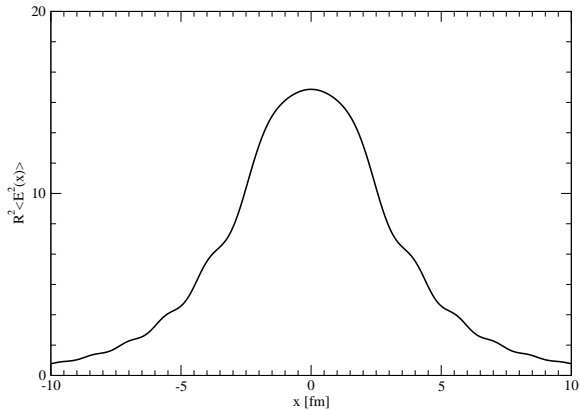}
 \caption{\label{fig1b} $R^2 \langle \E^2(x) \rangle$ in units of 
  $2b_C/\pi^3(\hbar c)^2 $ as a function of the distance $x$ transverse to the 
  $Q{\bar Q}$ axis. The units and the setting are as in Fig.~1.}
  \end{figure}

Finally, in Fig.~3, we show the contour plot of the energy
density as a function of the position in the $xz$ plane with quark
and antiquark on the $z$ axis at $R/2$ and $-R/2$ respectively. 
 \begin{figure}
 \includegraphics[width=3.in]{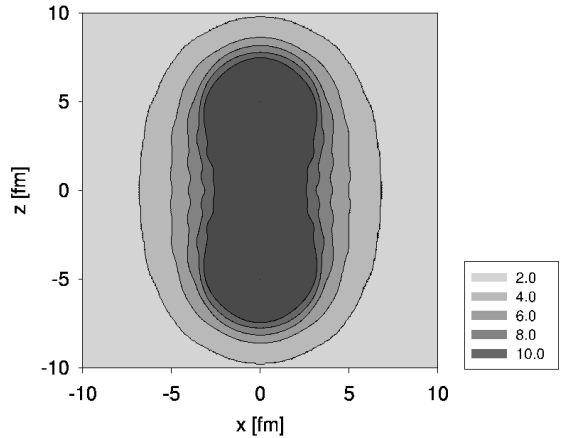}
 \caption{\label{fig2} $R^2 \langle \E^2(x) \rangle$ as a function of
    position in the $xz$ plane. The units and the same 
  setting as in Fig.~1. }
\end{figure}

It is clear from Figs.~2  and 3 that a flux tube like structure emerges and 
from \eref{short} that it has the correct scaling as a function of the 
$Q{\bar Q}$ separation but, as discussed above it does not have a finite 
correlation length (large $x$ behavior). 

The field distribution for the $QQQ$ system in the $f_L(p) =1 $
approximation is equal to the sum of three terms each representing
a contribution from a $QQ$ pair. We place each of the three quarks in a
corner of an equilateral triangle, $\z_i$, $i=1,\cdots3$
\begin{multline}
\langle \E^2(\x,\R_i) \rangle = \frac{C_F}{32\pi^2}
\bigl[ (\D_1 - \D_2)^2 \\ 
+ (\D_1 - \D_3)^2 + (\D_2 - \D_3)^2 \bigr],
\end{multline}
where 
\begin{equation}
\D_i = {{ \z_j - \x + \r/2 } \over 
 { |\z_j - \x + \r/2| }} d'_L(\z_i - \x - \r/2).
\end{equation}
 
The contour plot of of energy density in this case is shown in Fig.~4. 
Even though the field originates from the two-particle correlations the
net field seems to form into a ``Y''-shape structure.  This structure
has also recently been seen to emerge in Euclidean lattice simulations. 
 \begin{figure}
 \includegraphics[width=2.5in]{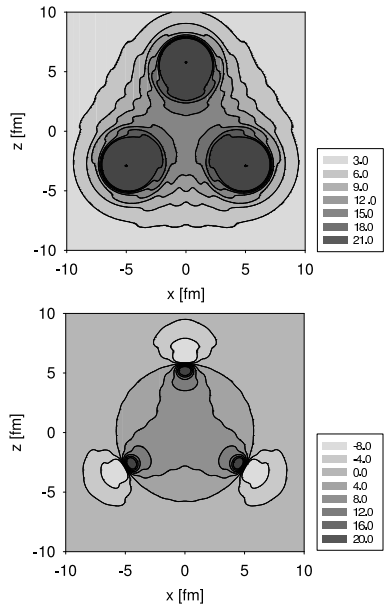}
 \caption{\label{fig3} $R^2 \langle \E^2(x) \rangle$ as a function of
    position in the $xz$ plane. The units and the same 
  setting as in Fig.~1. The upper panel show the total field
  distribution and the lower the distribution from mutual interaction
  (no self-energies) only. }
\end{figure}

 Finally, to study the effects of $f_L(p)$, in Fig.~5 we show the
 predictions for the $Q{\bar Q}$ field distribution 
 given by  \eref{vc} where we use $d(p)$ and $f(p)$ in the form given
  by  \eref{df} with  $\alpha = 1/2$ and $\beta=1$ and normalized such
   that $V(R) \to bR$ at large distances.  Furthermore, to remove
  the oscillations introduced by the momentum space cutoff,
  we
  now cut the small $x$ region in coordinate space, by i) extending
  the upper limits of integration in Eqs.~(\ref{dpl}) and~(\ref{fl}) to
  infinity and ii) cutting off the position space functions at short
  distances, 
 \begin{equation}
 d'_L(r) = {2\over {\pi\alpha}} \sin(\pi\alpha/2)
 \Gamma(2 -\alpha) \theta(r\mu - 1)
  { {\mu^2} \over {(\mu r)^{2-\alpha}}}  
 \end{equation}
 \begin{equation}
 f_L(r) = {1\over {2\pi^2}} \sin(\pi\beta/2)  
  \Gamma(2-\beta) \theta(r \mu - 1)
  { {\mu^3} \over {(\mu r)^{3-\beta}}}. 
 \end{equation}

 Comparing Fig.~3 and Fig.~5 we observe a narrowing of the flux tube.  
This is to be expected as the action of $f(p)$ is to introduce additional
gluonic correlations.  That said, there is no major qualitative change in the 
field distribution. 
 \begin{figure}
 \includegraphics[width=2.5in]{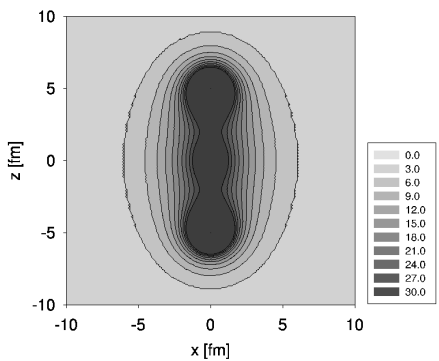}
 \caption{\label{fig4} $R^2 \langle \E^2(x) \rangle$ for $Q{\bar Q}$
   from \eref{el} with $\alpha = 1/2$ and $\beta=1$. 
  The units and the same setting as in Fig.~1, except that the contribution
  from $f(p)$ has been included.} 
\end{figure}

\section{Summary} 
We have calculated the distribution of the longitudinal chromo-electric
field in the presence of static $Q{\bar Q}$ and $QQQ$ sources using a
variational model for the ground state wave functional.  Despite
this wave functional having no string-like correlations a flux tube like
picture does emerge.  In particular the on-axis energy density
of the $Q{\bar Q}$ system behaves as $b_c/R^2$ for large inter-quark
separation, $R$ and the field falls off like $b_c R^2/x^4$ at large
distances from the center of mass of the $Q{\bar Q}$ system, $x$. This
 is weaker than in the Abelian case ($\sim R^2/x^6$) and implies that 
  moments of the average transverse spread of the tube, defined as 
 proportional to $\langle |x_\perp|^n \E^2(z,x_\perp) \rangle$, 
 are finite for $n<2$ only.  Thus there is no finite correlation length for 
 the longitudinal component of the chromo-electric field, as expected for the 
 state which does not take into account screening of the Coulomb line by the 
 transverse gluons (flux tube).  This also leads to large
 Van der Waals forces, which is bothersome, but it is consistent with
 the scenario of ``no confinement without Coulomb confinement'' of Zwanziger. 
 The Coulomb potential leads to a variational
 (stronger) upper bound to the true confining interaction. 
 
 Similar behavior at large distances is also true for the three quark
 sources, except that here we find the emergence of the ``Y''-shape junction.
 This is consistent with lattice simulations, but is remarkable in our case
 as it arises from two-body forces. It will be interesting to examine field
 distributions which include transverse field excitations. In that case
 the only lattice results available are for the potential, not 
 for the field distributions. Finally we note that, since the mean
 field calculation provides a variational upper bound, the long range
 behavior of the field distribution falls-off more slowly than expected for
 the Van der Waals force. Certainly as the complete string develops
 this is expected to disappear and it would be interesting to build a
 string-like model for the ansatz ground state to verify this
 assertion. 

\begin{acknowledgments}

The authors wish to thank J.\ Greensite, H.\ Reinhardt, Y.\ Simonov, 
F.\ Steffen, H.\ Suganuma and D.\ Zwanziger for helpful feedback.  This work 
was supported in part by the US Department of Energy under contract 
DE-FG0287ER40365.  The numerical computations were performed on the AVIDD 
Linux Clusters at Indiana University funded in part by the National Science 
Foundation under grant CDA-9601632.

\end{acknowledgments}

\end{document}